\documentclass[aps,prl,preprint,showpacs]{revtex4}  % for review and submission

\usepackage{graphicx}  % needed for figure
\usepackage{calc}  
\usepackage{dcolumn}   % needed for some tables
\usepackage{amsmath}

\usepackage[utf8]{inputenc}   % accents
\usepackage[T1]{fontenc}      % special caracters
\usepackage{textcomp}

\begin{document}
\title{Spatio-temporal wavefront shaping in a microwave cavity}
\author{Philipp del Hougne, Fabrice Lemoult, Mathias Fink, Geoffroy Lerosey}
\email[]{geoffroy.lerosey@espci.fr}
\affiliation{Institut Langevin, CNRS UMR 7587, ESPCI Paris, PSL Research University, 1 rue Jussieu, 75005 Paris, France}
\date{\today}

\begin{abstract}
Controlling waves in complex media has become a major topic of interest, notably through the concepts of time reversal and wavefront shaping. Recently, it was shown that spatial light modulators can counter-intuitively focus waves both in space and time through multiple scattering media when illuminated with optical pulses. In this letter we transpose the concept to a microwave cavity using flat arrays of electronically tunable resonators. We prove that maximizing the Green's function between two antennas at a chosen time yields diffraction limited spatio-temporal focusing. Then, changing the photons' dwell time inside the cavity, we modify the relative distribution of the spatial and temporal degrees of freedom (DoF), and we demonstrate that it has no impact on the field enhancement: wavefront shaping makes use of all available DoF, irrespective of their spatial or temporal nature. Our results prove that wavefront shaping using simple electronically reconfigurable arrays of reflectors is a viable approach to the spatio-temporal control of microwaves, with potential applications in medical imaging, therapy, telecommunications, radar or sensing. They also offer new fundamental insights regarding the coupling of spatial and temporal DoF in complex media.
\end{abstract}

\pacs{42.25.Dd, 41.20.Jb, 84.40.-x}
\maketitle

Wave propagation in complex media is known to cause a complete scrambling of the input wavefronts, due to multiple scattering or reverberation of waves. As a result, the wave fields in these media resemble optical speckles both in space and in time \cite{goodman_speckle,NatPhotReview}. Yet information transfer through complex media is crucial for many applications in telecommunications, imaging or medical therapies. Examples of a complex medium at optical frequencies include highly scattering opaque ones as well as biological tissues or multimode fibers \cite{choi2015WSforBioMed,park2013WSforOCT,choi_biomed,pap_biomed,bianchi_biomed,cizmar_biomed}; in the microwave domain, forests or cities can be considered multiple scattering media, while reverberating media are also very common, ranging from reverberation chambers for electromagnetic compatibility tests, via open disordered cavities for computational imaging to indoor environments trapping wireless communication signals \cite{hill_electromagnetic_2009,DavidSmith_CompImag_APL,PhaselessCompImag_DavidSmith,SMM_PoC}. Numerous techniques, notably time reversal and wave front shaping \cite{TR_fink,EMTR_prl,mosk_SLM,Mosk_WS_OptTrans,EigChan_Choi_NPhot,BretagnePRE,Anlage_syntheticTR,Anlage_ExpCompTR,Anlage_NLlossyTR,kuhl_1DWS,publikation1,Gigan_NonInvImagPR,Gigan_NonInvTM_PA,LD_binDMD,TM_RevMed}, have been proposed to take advantage of the multiple scattering and reveberation occuring during propagation. A common ground of these approaches is that they make use of the secondary sources offered by scatterers and reflectors, which provide additional degrees of freedom (DoF), to the point that they can even outperform focusing in homogeneous media \cite{EMTR_science,park2013subwavelength,mosk_disorder4perfectFOC,choi2011subwavelengthFOC,Mosk_vis_subwavelength_foc_byWS}.

Time reversal, which is naturally a broadband approach, results in spatio-temporal focusing of waves. As a consequence, it provides a maximum of acoustic or electromagnetic intensity at a given time and a given location that may be employed for various applications such as medical therapy, electronic warfare and wireless or underwater communications \cite{TR4Lithotripsy,acoustic_bazooka,EM_bazooka,kuperman_oceanTR,Kuperman_TR4underwaterComm,Geof_TR4comm,naqvi_TRcomm}. On the other hand, wavefront shaping, because it acts in the spatial domain, is originally a monochromatic concept that results in maxima of deposited energy at desired foci. Nevertheless, the mixing of spatial and temporal DoF in complex media was studied and exploited in acoustics, to use temporal DoF for spatial focusing \cite{fab_ST}. Similarly, exploiting the fact that a spatial control over a transient wave field can also act on the time dependence of the transmission through a complex medium, demonstrations of spatio-temporal focusing by wavefront shaping were recently reported in optics, initially based on closed-loop iterative optimization schemes and more recently also based on open-loop transmission matrix approaches \cite{aulbach_STF,katz_STF,gigan_STF,dariaSciRep,mikael_STF}. 

% Set-up & Method Figure
% [FIGURE HAS TO APPEAR IN LATEX TEXT BEFORE THE FIRST PARAGRAPH THAT (PARTIALLY) APPEARS ON PAGE 2]
\begin{figure*}[bt]
	\begin{center}
\includegraphics [width=\textwidth] {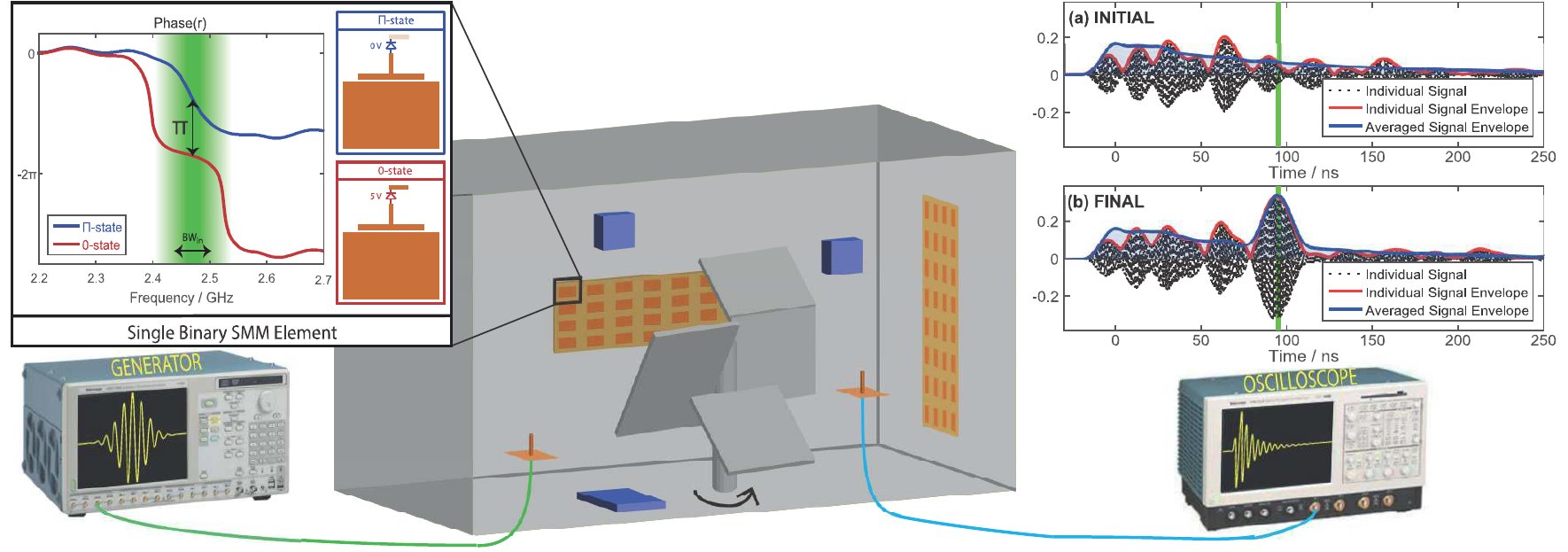}
	\caption{Schematic of the experimental set-up, displaying two monopole antennas placed in a reverberant disordered cavity, a spatial microwave modulator (SMM), a mode-stirrer and some electromagnetic absorbers (blue). The emitting antenna is fed by an arbitrary signal generator, while the receiving one is connected to a high sampling rate oscilloscope. The inset (adapted from \cite{SMM_design}) illustrates the characteristics of a two-state single SMM element. (a) and (b) show exemplary how, using the SMM, we are able to focus the cavity's Green's function envelope at a chosen optimization time $t_{\mathrm{opt}}$ (indicated in green). The optimum SMM configuration is identified iteratively with a continuous sequential algorithm.}
	\label{fig1}
	\end{center}
\end{figure*}

In this letter, we transpose this concept back to the microwave domain, using spatial microwave modulators (SMMs), that is, arrays of resonators that are electronically reconfigurable with simple logical controls \cite{SMM_design}. To do so, we work in a reverberant cavity, whose surface has been partly covered by a SMM. We first prove that shaping the wave field inside the cavity in the time domain, by maximizing the envelope of the transient Green's function between two antennas at a chosen time, results in spatio-temporally focused microwaves. We then present a parametric investigation of the enhancement in spatio-temporal focusing under well-controlled conditions. The choice of a microwave cavity as a complex medium ensures a fixed volume and different dwell times are easily explored by changing the cavity's quality factor $Q$. As the latter governs the distribution of the cavity's DoF between space and time, this system is a good candidate to accurately study the interplay between spatial and temporal effects. We examine the characteristics in time and space of spatio-temporal focusing for different cavity dwell times and subsequently quantify both the instantaneous signal enhancement as well as the total energy deposited at the target position by spatio-temporal wavefront shaping.

The typical experiment, illustrated schematically in Fig.~1, consists in measuring the transmission between two monopole antennas placed in a metallic reverberant cavity that is disordered (volume $1.1 \ \mathrm{m^3}$, surface area $6.6 \ \mathrm{m^2}$). One emits a pulse centered on $f_0=2.47 \ \mathrm{GHz}$ with a bandwidth of $\Delta f_{in} = 66 \ \mathrm{MHz}$, chosen to be smaller than that of the SMM. Using the other one, we record the corresponding transient Green's function with a sampling rate oscilloscope. It contains the ballistic coherent signal, originating from the direct path, and a long exponentially decaying coda originating from multiple reverberations, whose typical duration $\tau$ equals the dwell time of the photons inside the cavity.

To achieve our objective of maximizing the envelope of the received signal at a desired time, exemplary indicated by the vertical green line in Fig.~1(a,b), we cover $7\%$ of the cavity walls with our SMM, consisting of $102$ elements whose dimensions are on the order of $\lambda_0 /2$. Each element is electronically controllable to be in either of the two states explained in the inset of Fig.~1 \cite{SMM_design}. At and around $f_0$, there is a difference of $\pi$ in the phase of the respective reflection coefficients. By modifying the SMM configuration, we change the distribution of times of pulse arrivals and hence the shape of the Green's function. Iteratively we identify step by step the best configuration to satisfy our objective, using the feedback obtained from the oscilloscope \cite{moskWSalgo}.

As exemplary shown in Fig.~1(a,b), with the above setup and procedure we can focus the transient electromagnetic energy at a desired time. The iterative procedure eventually identifies a SMM configuration that matches the phases of different paths such that they interfere constructively at this time. When working with complex media, a single realization is yet usually not at all representative of the system's average behavior; one might for instance start from a minimum, resulting in an extraordinary enhancement. To obtain statistically significant results, many realizations over disorder are required. We conveniently achieve this by rotating the mode-stirrer shown in Fig.~1 in steps of $12^{\circ}$, which results in a completely uncorrelated ``new'' cavity; and we repeat the experiment for two independent positions of the emitter. This allows us to average the results of our experiments over up to $90$ realizations of the disordered cavity.

% TEMPORAL FIGURE
\begin{figure}[bt]
	\begin{center}
\includegraphics [width=8.6cm] {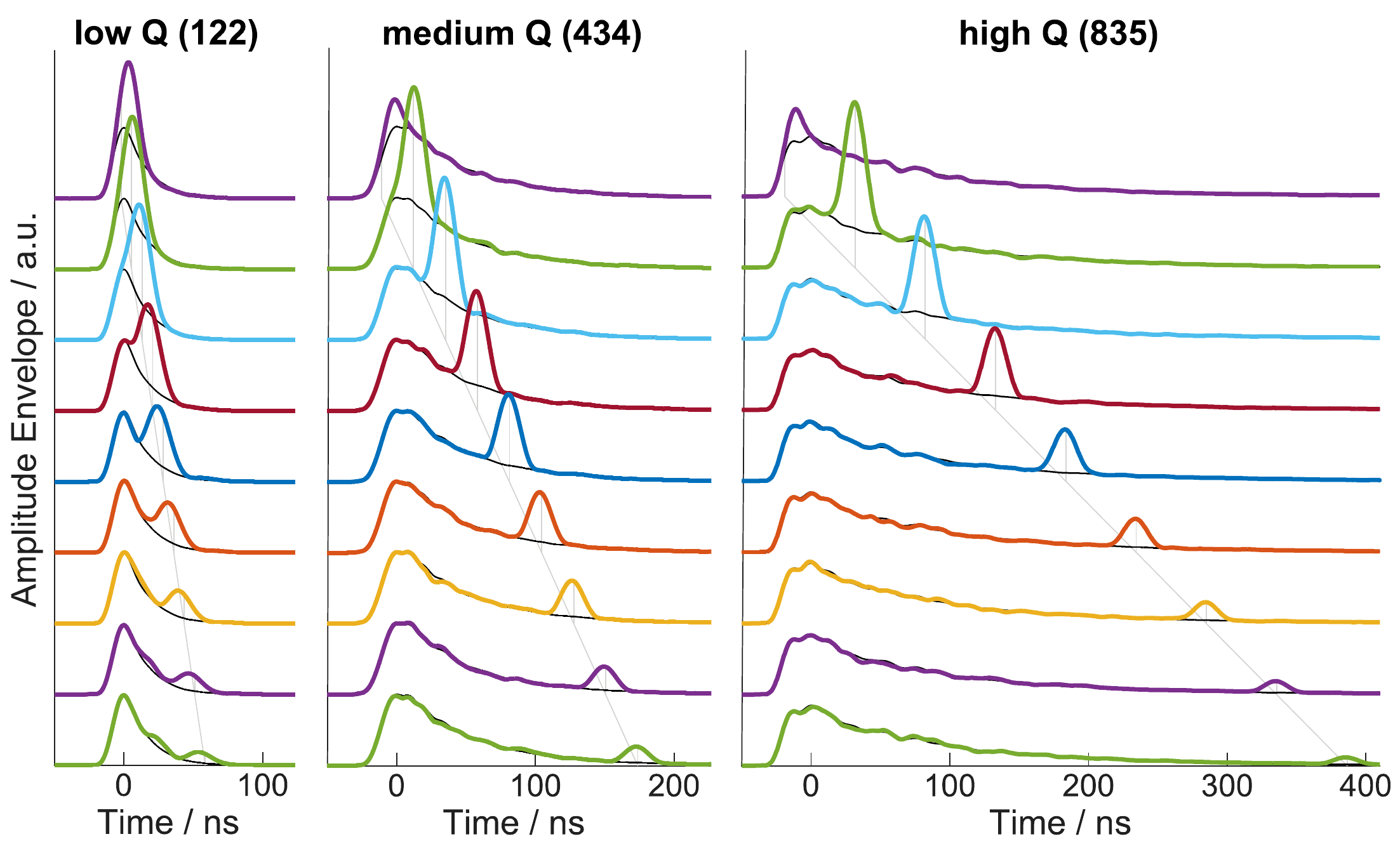}
	\caption{Green's function envelopes averaged over 60 realizations of the disordered cavities, before (thin black line) and after (thick color lines) temporal focusing of the waves using the SMM and the iterative algorithm on a set of chosen times. The series of experiments has been realized for 3 different quality factors $Q$ of the cavity. The duration of the focused peaks is always the same: that of the emitted pulse.}
	\label{fig2}
	\end{center}
\end{figure}

To comparatively investigate different cavity dwell times $\tau$, electromagnetic absorbers are glued to the cavity walls (in small pieces, distributed approximately isotropically) to control the independent variable $Q$. For three different quality factors, we display in Fig.~2 averaged Green's function envelopes before and after optimization, for a selection of optimization times. A clear temporal focusing effect is visible, at any time of the Green's functions between the two antennas. We observe that the duration of the optimized pulse is $15 \ \mathrm{ns}$, depending neither on the chosen optimization time nor on the cavity dwell time. Similarly to time reversal experiments, we hence verify here that the temporal length of the focused wave field is of the order of that of the pulse emitted by the source antenna.

The question that arises now is: what happens spatially?
To explore this, we probe the field spatially by replacing the receiving antenna with a line of antennas, separated by about $\lambda_0/10$, and of reduced length to minimize near-field coupling. Displacing a single antenna in the cavity is not feasible as it would invasively impact on the cavity's boundary conditions. Fig.~3 illustrates averaged focusing examples (similar curves as in Fig.~2 but simultaneously probed off the objective antenna) for all three values of $Q$.  Those maps exhibit again the time focusing, but surprisingly we note that the energy enhancement is also concentrated spatially around the target antenna. The temporal focusing has thus simultaneously carried out spatial focusing, with a focal width which reaches the diffraction limit of $\lambda_0/2$. 

This effect, even if it appears counter-intuitive, is actually a direct consequence of the speckle-like nature of the wave field within a disordered cavity. Each frequency content excited within the cavity when the antenna emits a pulse is also spatially distributed onto speckle grains. And, in a cavity, because the boundaries isotropically surround the receiving antennas, their spatial extent is half a wavelength. The SMM allows to match the phases of the uncorrelated speckle grains at the desired position but acts randomly at every other uncorrelated position, overall resulting in the spatial focusing. Note that we have also verified that similar spatio-temporal maps are obtained whatever the optimization time, thus guaranteeing that all of the temporal results presented in Fig.~2 gave rise as well to a spatial focus onto $\lambda_0/2$ wide spots. This has a profound interest in terms of applications since one only needs to record the wave field at a given position and a single time to ensure that the waves are also spatially focused thanks to the complex nature of the propagation medium.

% SPATIAL FIGURE
\begin{figure}[bt]
	\begin{center}
\includegraphics [width=8.6cm] {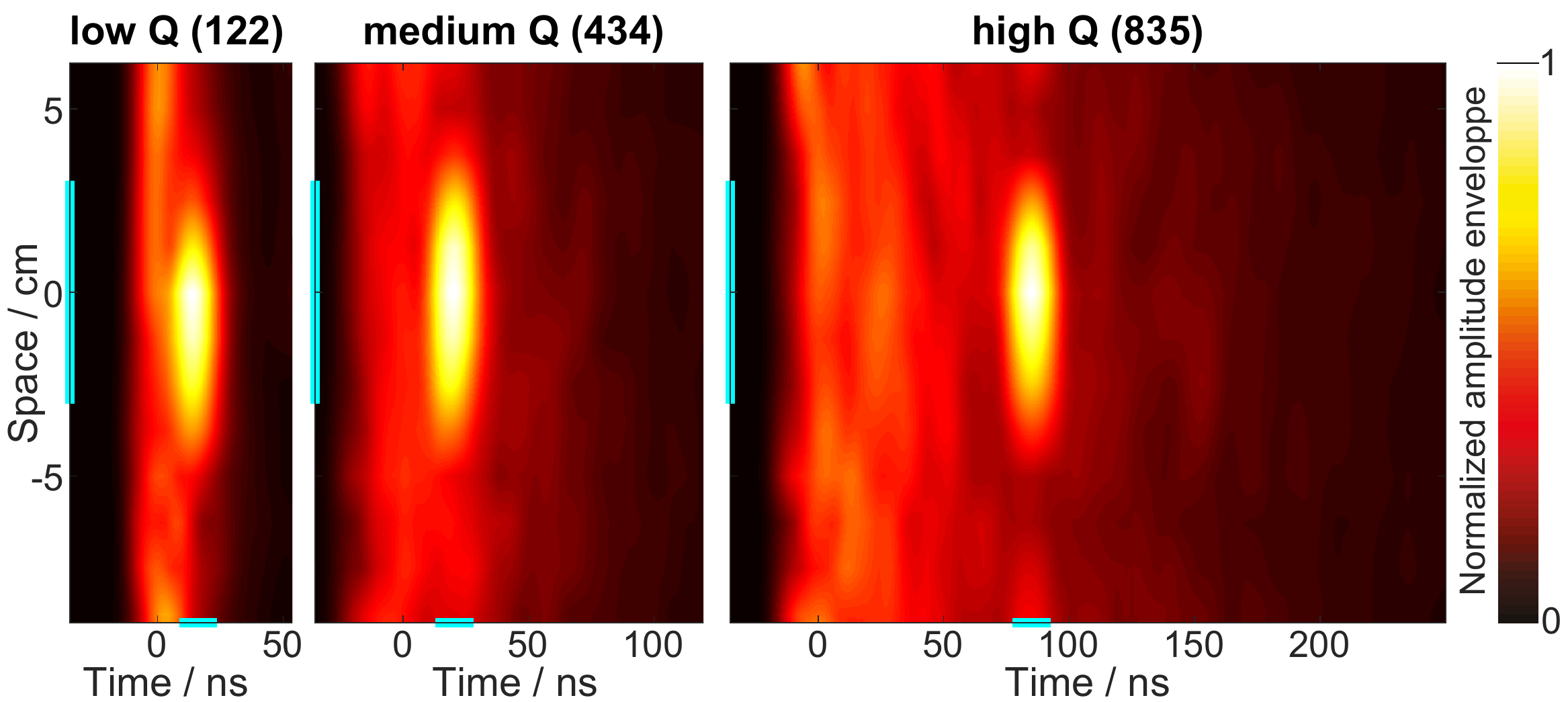}
	\caption{Spatio-temporal maps of the Green's functions envelopes around a given time and position chosen for wavefront shaping, for three different quality factors, and averaged over 30 realizations of disorder. Each temporal wavefront shaping experiment results in a spatio-temporally focused wave field, onto spots of dimensions $\lambda_0 /2$ in space and $1/\Delta f_{in}$ in time.}
	\label{fig3}
	\end{center}
\end{figure}

Having characterized the effect of spatio-temporal focusing in space and time, we now quantify the attainable enhancements for different dwell times $\tau$ of the photons in the cavities. Firstly, we consider what we define as instantaneous spatio-temporal enhancement $\eta_A$ at the chosen optimization time $t_{\mathrm{opt}}$:
\begin{equation}\label{eqnEta}
	\eta_A = \frac{\langle h_{\mathrm{fin}}(t_{\mathrm{opt}}) \rangle}{\langle h_{\mathrm{init}}(t_{\mathrm{opt}}) \rangle},
\end{equation}
where $h$ is the measured Green's function envelope. In Fig.~4(a) we show the average instantaneous enhancements $\eta_A$ achieved at different optimization times, for the three different values of $Q$. Note that the curves are displayed in time units normalized by $\tau$, for ease of comparison. All three curves superpose and a plateau, for which $\eta_A=6$, is obtained in all three cases after the ballistic signal but before attenuation becomes too important.
 
% FIG 4
\begin{figure}[bt]
	\begin{center}
\includegraphics [width=8.6cm] {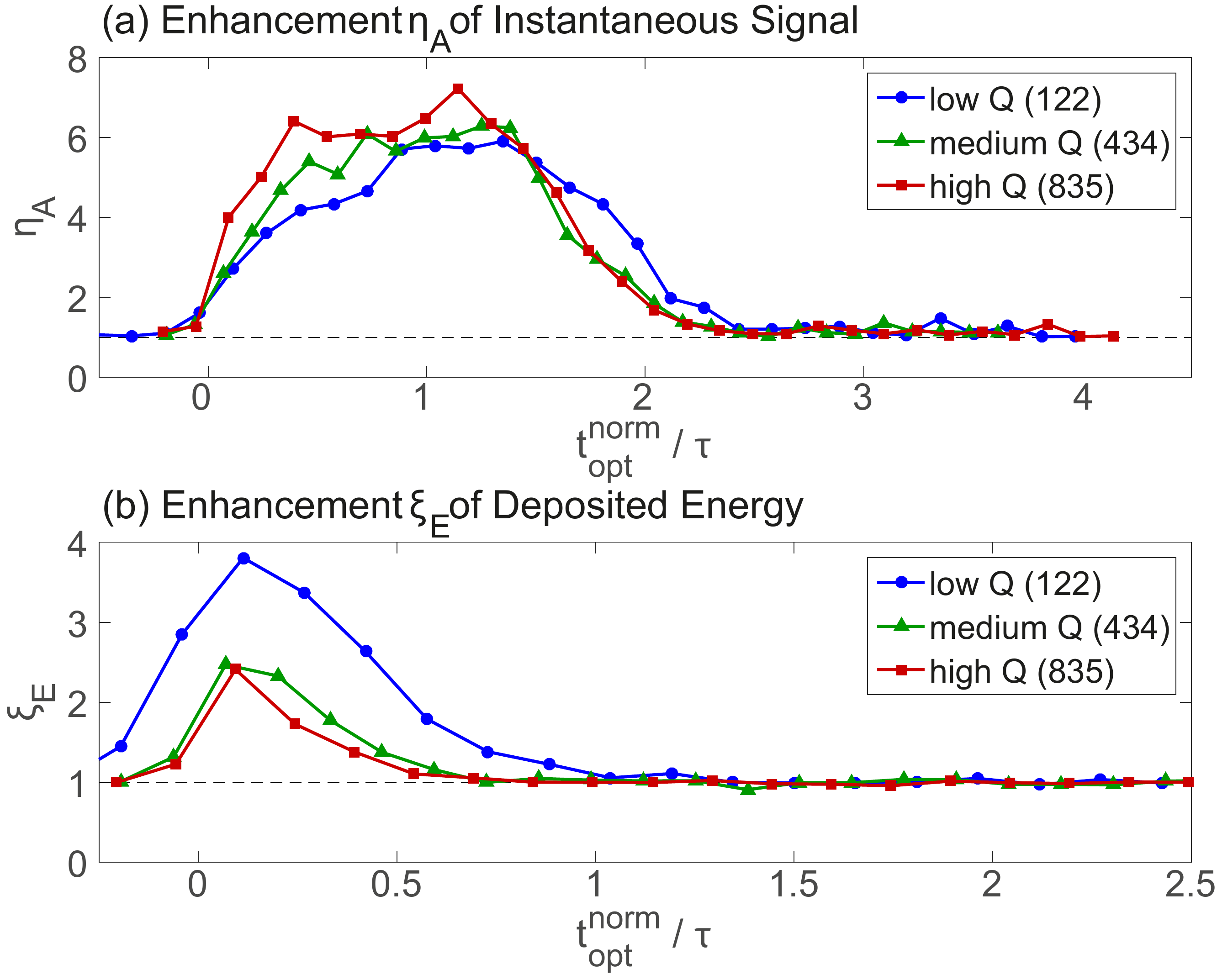}
	\caption{Quantification of the impact of different cavity dwell times $\tau$ on the enhancement obtained by spatio-temporal wavefront shaping at different optimization times $t_{\mathrm{opt}}^{\mathrm{norm}}$ (for clarity in units of the cavity dwell time $\tau$ corresponding to each quality factor). (a) Instantaneous spatio-temporal enhancement $\eta_A$. (b) Deposited energy enhancement $\xi_E$. All results are based on 60 independent realizations of the disordered cavity.}
  \label{fig4}
	\end{center}
\end{figure}

At this point, it is instructive to evaluate the number of spatial $N_S$ and temporal $N_T$ DoF that participate in the spatio-temporal focusing achieved for various $Q$ of the cavity. Analogous to time reversal experiments, the spatio-temporal focus is indeed the coherent sum of $N_T$ uncorrelated frequency components, each of which results from the coherent sum of $N_S$ independent overlapping eigenmodes of the cavity within a correlation frequency $f_{corr}$ \cite{fab_ST,derodePRE1,derodePRE2,publikation1}. 

The number of spatial DoF within a correlation frequency is given by \cite{publikation1} \footnote{Note that we consider cases where $\Delta f_{in} \ll f_0$; hence our estimate of $N_S$ made for the central frequency $f_0$ is appropriate for any frequency within the considered bandwidth.}:
\begin{equation}
	N_S = 4 \pi V \frac{f_0^2}{c^3} \times f_{corr} = 4 \pi V \frac{f_0^3}{c^3} \times \frac{1}{Q}.
\end{equation}
The amount of temporal DoF is simply the number of statistically independent frequencies within the bandwidth $\Delta f_{in}$: 
\begin{equation}
	N_T = \frac{\tau}{\Delta t_{in}} = \frac{\Delta f_{in}}{f_{corr}} = \frac{\Delta f_{in}}{f_0} \times Q. 
\end{equation}
Hence, the total number of spatio-temporal DoF can be evaluated as:
\begin{equation}
	N = N_S \times N_T =  4 \pi V \frac{f_0^2}{c^3} \times \Delta f_{in}.
\end{equation}

It is important to note that, for each experiment, the total number of DoF is the same, being independent of the cavity's quality factor $Q$. Yet we clearly see in these formulae that $Q$ controls the repartition of the system's DoF between space and time. As a consequence, since the average behavior of the instantaneous spatio-temporal enhancement $\eta_A$ is the same for the three cases with very different distributions of DoF, we conclude from our results that spatio-temporal wavefront shaping tends to use all DoF available in the system, whether they are of temporal or of spatial nature. Stated differently, in a problem where there are few uncorrelated frequencies available, the operation will tend to focus waves spatially similarly to a monochromatic experiment, while with a large number of independent frequencies in the bandwidth, wavefront shaping will mainly result in the synchronization of these frequencies at a chosen time.

This can be underlined by quantifying how the total amount of energy received at the target position is affected by the spatio-temporal focusing. Indeed, by considering the ratio between the deposited energy after and before optimization:
\begin{equation}\label{eqnXi}
	\xi_E = \frac{\langle \int \! h^{2}_{\mathrm{fin}}(t) \, \mathrm{d}t \rangle}{\langle \int \! h^{2}_{\mathrm{init}}(t) \, \mathrm{d}t \rangle}.
\end{equation}
This removes any temporal coherent effect. We are left only with the focusing enhancement in space, hence revealing to what extent the spatio-temporal focusing profited from focusing in time.

We show in Fig.~4(b) the average energy enhancements $\xi_E$ achieved at different optimization times, for the three different values of $Q$. Now, the quality factor matters: the cavity with the lowest dwell time achieves the highest values of $\xi_E$. The enhancement of the total energy deposited at the target position by wavefront shaping thus decreases as a higher proportion of DoF is of temporal nature. This last comment allows us to point out a clear difference between time reversal and spatio-temporal wavefront shaping. Indeed, when using time reversal to focus waves in space and time, a long transient Green's function is recorded and sent back into the medium. This generates very energetic foci, because a very long time-varying wave-field is compressed in time at the collapse time. Hence, longer dwell times of the photons inside the cavity result in higher deposited energies. This is not the case using wavefront shaping since, in any case before and after shaping the wave field, the same pulse and consequently the same amount of energy entering the medium was emitted. We finally note that total energy and spontaneous enhancements have very different temporal shapes; this is easily understandable since, when considered the total energy deposited, optimizing at early times within the Green's function has a much bigger effect than at late times. 

To conclude, in this work we have introduced the concept of spatio-temporal focusing by wavefront shaping in the microwave domain, in a cavity. We have parametrically investigated the influence of the distribution of the degrees of freedom between space and time. While the same instantaneous signal enhancement was observed for different cavity dwell times, the use of spatial DoF was more important the lower the cavity's quality factor was, leading to higher enhancements of the total energy delivered to the target. We believe that our set of experiments carried out in the microwave domain lends new insights into related works with complex media, notably in optics where a precise control of the independent variables of the problem might be difficult or unfeasible. Moreover, transposing the concept of spatio-temporal wavefront shaping from optics to the microwave range, using very simple electronically controllable SMMs, may offer a wealth of applications. These could be found in the domains of radars, antennas, high power sources, imaging devices, or wireless communications for instance \cite{acoustic_bazooka,EM_bazooka,DavidSmith_CompImag_APL,PhaselessCompImag_DavidSmith,Geof_TR4comm,naqvi_TRcomm}.

P.d.H. acknowledges funding from the French ``Ministère de la Défense, Direction Générale de l’Armement''. This work is supported by LABEX WIFI (Laboratory of Excellence within the French Program ``Investments for the Future'') under references ANR-10-LABX-24 and ANR-10-IDEX-0001-02 PSL* and by Agence Nationale de la Recherche under reference ANR-13-JS09-0001-01.

%\bibliography{references_stf}

\end{document}